\documentstyle[prl,preprint,epsfig,aps]{revtex}
\draft
\tighten
\renewcommand{\vec}[1]{\mbox{\boldmath$#1$\unboldmath}}
\begin{document}

\title{Structure of the exotic nucleus $^{14}$B in the ground state}
\author{R. Chatterjee and P. Banerjee}
\address{Theory Group, Saha Institute of Nuclear Physics,
\\1/AF Bidhan Nagar, Calcutta - 700 064, INDIA}
\date{\today}
\maketitle

\begin{abstract}
We investigate the structure of the neutron rich nucleus $^{14}$B
through studies of its breakup in the Coulomb field of a heavy target.
The breakup amplitude is calculated within an adiabatic as well as
finite range DWBA theories of breakup reactions. 
Both these formalisms allow the use of realistic
wave functions for the relative motion between the fragments in the
ground state of the projectile. The longitudinal momentum distributions 
of $^{13}$B (ground state) following the breakup of
$^{14}$B on a heavy target at beam energy of 60 MeV/nucleon,
calculated using two possible ground state configurations of $^{14}$B, 
have been compared with the recent data. The data seem to favour 
$^{13}$B(${3\over 2}^-)\otimes$2$s_{1/2}$ as the possible ground state
configuration of $^{14}$B with a spectroscopic factor close to unity. 
We give our predictions for the neutron angular distributions and
the relative energy spectra in the one-neutron removal reaction of $^{14}$B.
\end{abstract}


It has now been well established that close to the neutron drip line, 
one encounters nuclei which have a long tail of one or two neutrons  
outside a nuclear core -- a feature commonly known as a neutron halo
\cite{han}. It has been observed that nuclei with a dominant configuration 
of loosely bound $s$-wave valence neutron(s) generally have a well-developed 
halo structure.
Breakup reactions, in which the halo particle is removed from the
system, are promising tools in investigating the structure of these
nuclei. Thus, halo formation in such nuclei can be investigated by measuring 
the longitudinal or parallel momentum distribution (PMD) of fragments from the
breakup reaction \cite{baz}. The wide spatial dispersion characteristic
of a halo neutron translates into a narrow momentum distribution. 

Early experiments on breakup reactions were based on the assumption that the 
measured momentum
distribution represented a single reaction channel. It has recently become 
possible to go beyond this simple approach by measuring the $\gamma$-radiation 
from the decay of excited states of the core in coincidence with the breakup 
fragments \cite{nav}. Thus breakup events, in which the core remains in the
ground state, are identified by associating them with measured $\gamma$-ray 
multiplicity equal to zero. 
This technique, therefore, allows for spectroscopic investigation of both the
core nucleus and the valence nucleon, by measuring the partial cross
sections for the excitation of different final states of the core.

Among the candidates for neutron halo nuclei, the $^{14}$B nucleus (with
valence neutron separation energy of 0.97 MeV) is
of particular interest as it is the lowest mass bound system among the $N$= 9
isotones. The well-known halo nuclei $^{11}$Be, $^{11}$Li, $^{14}$Be, $^{17}$B
and $^{19}$C occupy a similar position for $N$= 7, $N$= 8, $N$= 10, $N$= 12 and
$N$= 13 respectively. Moreover, $^{14}$B is an odd-odd nucleus and thus 
different from any other neutron halo system observed to date.

Moderate enhancements have been observed in the total reaction cross section
measurements for $^{14}$B at intermediate energies \cite{sai,vil}, thereby
giving it a large root mean square (rms) radius. Non-linear relativistic mean field 
calculations predict inversion of 1$d_{5/2}$ and 2$s_{1/2}$ orbitals (as in
the case of the well-established one-neutron halo nucleus $^{11}$Be) in the
ground state of $^{14}$B \cite{ren}. Very recent measurement at GANIL on the 
core PMD
resulting from the one-neutron removal reaction of $^{14}$B on a carbon
target at 50 MeV/nucleon incident energy gives a narrow width (FWHM = 56.5
$\pm$ 0.5 MeV/c) \cite{sau}, which is a characteristic of the halo structure. 
All these facts indicate
the possibility of the presence of an extended neutron distribution in this
nucleus, although the one-neutron separation energy of almost 1 MeV is
likely to suppress the development of as large a distribution as that found
in the more weakly bound one-neutron halo nuclei $^{11}$Be and $^{19}$C.
It should, however, be noted that $^{15}$C has a valence neutron 
separation energy of $\simeq$1.2 MeV, which is even larger than that
of $^{14}$B. But it is most likely a one-neutron halo nucleus \cite{chatt},
because the valence neutron has a dominant $s$-state configuration in the
ground state of $^{15}$C. On the other hand, although the one-neutron 
separation energy in case of $^{17}$C is only 0.729 MeV, 
the halo formation in this nucleus is hindered because 
of a strong $d$-wave configuration of the valence neutron in its ground
state \cite{baz,chatt}.

Recently the PMD's of the $^{13}$B core fragment have been measured in 
one-neutron knock-out reactions from $^{14}$B on both light ($^9$Be) and heavy 
($^{197}$Au) targets at around 60 MeV/nucleon beam energy at MSU \cite{gui}. 
Calculations on the parallel momentum distributions, when $^{13}$B remains in 
its ground state, have been compared with the data for both the targets 
\cite{gui}. The agreement between the calculated width and experimental width 
(59$\pm$3 MeV/c) has been found to be consistent with expectations for a 
dominant weakly bound 2$s_{1/2}$ neutron configuration, which lends 
support to the existence of a one-neutron halo structure in $^{14}$B. 

The mechanism of the breakup reaction of the loosely bound exotic nuclei on 
heavy targets is generally known to be dominated by Coulomb dissociation, 
particularly below the grazing angle. 
The grazing angle for the MSU reaction on Au target is around 4.4$^\circ$.
The angular acceptances for detection of the charged $^{13}$B core were 
$\pm$3.5$^\circ$ and $\pm$5$^\circ$ in the dispersive and non-dispersive
directions. 
%i.e. in two mutually perpendicular directions in the reaction plane.
Therefore, it is expected that the heavy target data have mostly contributions 
from the Coulomb dissociation.

Using a Yukawa potential with finite-size corrections for the core-neutron
relative motion wave function, the authors of Ref.~\cite{gui} performed 
semiclassical calculations for Coulomb breakup of $^{14}$B on the Au target and 
compared their results with the data for the events in which the core remains
in the ground state. The total one-neutron
removal cross section of 543 mb calculated by them is somewhat less than the 
measured cross section of 638$\pm$45 mb. The calculated PMD had to be 
normalized (by a factor greater than unity) to the experimental data in order 
to get a fit \cite{gui}. However, the agreement in the wings of the distribution 
was still unsatisfactory.
The zero-range calculations done in Ref.~\cite{gui} also does not allow one
to draw inference on spectroscopic factors of the possible configurations
present in the ground state of $^{14}$B, as far as the data on the heavy target
is concerned.

In this Brief Report, we perform Coulomb breakup calculations on the PMD of
$^{13}$B (ground state) resulting from the breakup of $^{14}$B on Au and compare
the results with the MSU data with a view to put constraint on the spectroscopic
informations of the probable configurations in the ground state of $^{14}$B. 
We use two distinct theoretical approaches - the adiabatic (AD) model 
\cite{tos,ban,ban99} and the finite range (post form) distorted wave Born approximation 
(FRDWBA) method \cite{chatt}. They give 
almost the same expression for the reaction amplitude, but differ in details
of the formulation. The adiabatic theory of the Coulomb breakup of a projectile
used by us has been described extensively in \cite{tos,ban,ban99}. The triple
differential cross section for the reaction, in which a projectile $a$ 
breaks up into a charged core $c$ and a neutral valence particle $n$ on 
target $t$, is given by (assuming an inert core)
\begin{eqnarray}
{d^3\sigma \over dE_c d\Omega _cd\Omega _n}={2\pi\over \hbar v_a}\left\{\sum_{l
\mu}\frac{1}{(2l + 1)}\vert \beta^{\rm AD}_{l\mu}\vert^2 \right\} \rho(E_c,
\Omega _c,\Omega _n)~.
\end{eqnarray}
Here $v_a$ is the $a$--$t$ relative velocity in the entrance channel and
$\rho (E_c,\Omega _c,\Omega _n)$ the phase space factor appropriate 
to the three-body final state. The reduced amplitude $\beta^{\rm AD}_{l\mu}$
is given by
\begin{eqnarray}
\beta^{\rm AD}_{l\mu}
=\langle \vec{q}_n\vert V_{cn}\vert \Phi _{a}^{l\mu}\rangle
\langle \chi ^{(-)}(\vec{k}_c);\alpha\vec{k}_n\vert \chi
^{(+)}(\vec{k}_a) \rangle~.
\end{eqnarray}
The first term in Eq. (2) contains the structure information about
the projectile through the ground state wave function 
$\Phi _{a}^{l\mu}(\vec{r})$, and it is
known as the vertex function \cite{ban}, while the second
term is associated only with the dynamics of the 
reaction, which can be expressed in terms of the
bremsstrahlung integral \cite{nor}. For explanation of other quantities 
in the above equation, we refer to \cite{ban}. It should, however, be mentioned
that to obtain Eq. (2) \cite{tos}, it has been assumed that the dominant
projectile breakup configurations excited are in the low-energy continuum
(the adiabatic approximation). 

For the FRDWBA case, the triple differential cross section looks the same
excepting the reduced amplitude, which is given by
\begin{eqnarray}
\beta^{\rm FRDWBA}_{l\mu}
=\langle (\gamma\vec{q}_n-\alpha\vec{K})\vert V_{cn}\vert \Phi _{a}^{l\mu}
\rangle \langle \chi ^{(-)}(\vec{k}_c);\delta\vec{k}_n\vert \chi
^{(+)}(\vec{k}_a) \rangle~.
\end{eqnarray}
In the above, $\vec{K}$ is an effective local momentum associated with the
core-target relative system, whose direction has been taken to be the same as 
the direction of the asympotic momentum $\vec{q}_c$ \cite{chatt}. For 
clarifications about other quantities, we refer to \cite{chatt}.

Both the theories are fully quantum mechanical. The adiabatic formalism 
is non-perturbative, while the FRDWBA formalism assumes that the breakup
states are weakly coupled. Both
the methods retain finite-range effects associated with the interaction
between the breakup fragments and include the initial and final state
Coulomb interactions to all orders. The theories allow the use of wave 
functions of any relative orbital angular momentum $l$ for the motion between
$c$ and $n$ in the ground state of $a$. 

The spin-parity of $^{14}$B is known to be 2$^-$ in its ground state.
Shell model calculations with the WBP and WBT residual interactions suggest 
spectroscopic factors of 0.306 and 0.662 respectively for the removal of a
valence neutron from the 1$d_{5/2}$ and 2$s_{1/2}$ states in $^{14}$B 
\cite{war}. For our calculations, we have used a single particle
potential model for $^{14}$B in which the valence neutron, with a binding
energy of 0.97 MeV, moves relative to the $^{13}$B core (with intrinsic 
spin-parity ${3\over 2}^-$) in a Woods-Saxon
potential with radius and diffuseness parameters 1.15 fm and 0.5 fm
respectively. The depths of this potential well have been adjusted to reproduce
the valence neutron binding energy. We have considered 
two possible configurations for the valence neutron in the $^{14}$B
ground state: (a) a 2$s_ {1/2}$ state bound to the $^{13}$B core
and (b) a 1$d_{5/2}$ state bound to the $^{13}$B core. 
The rms sizes of $^{14}$B with options (a) and (b) were found to be 
2.79 fm and 2.57 fm respectively, while the 
corresponding rms sizes of the valence neutron were 5.46 fm
and 3.43 fm respectively. The rms size used for the $^{13}$B 
core is 2.5 fm \cite{baz0}. 

In Fig. 1, we have compared our calculations with the data for the PMD of
$^{13}$B (ground state) in the breakup of $^{14}$B on Au at 60 MeV/nucleon
beam energy. In these calculations, the maximum values of the core 
transverse momentum integrations correspond to the 
maximum angles of detection of the core as in the MSU experiment \cite{gui}.
Calculations corresponding to pure $s-$state and pure $d-$state
configurations of the valence neutron are shown by the solid and dashed lines
respectively, while a coherent superposition of these two results weighted by
the spectroscopic factors of Ref.~\cite{war} is given by the long-dashed curve.
The AD model results are given by thick lines, whereas those of
FRDWBA by thin lines. This same convention has been followed
in case of subsequent figures also. We see that both the theories give
similar results.
The pure $d$-state contribution is much less than the pure $s$-state one.
The calculated width for the pure $s-$state result is about 95\% of the 
measured width of 59$\pm$3 MeV/c, with the experimental error of $\pm$5\%
\cite{kol}. It gives almost 80\% of the total one-neutron removal cross section,
which is 510.4 mb\cite{kol}. But the fit to the experimental data worsens
if we consider a non-negligible $l=$2 admixture in the wave function
of $^{14}$B (long-dashed curves), taking the spectroscopic factors of the 
$s$-state and $d$-state neutrons as in Ref.~\cite{war}. In any case, the theoretical 
cross section is less than the experimental one.

The binding energy of the valence neutron(s) plays a crucial role in determining
the relative importance of Coulomb and nuclear breakup cross sections of the
halo nuclei in different kinematical domains. It has been found in case of 
$^6$He, with a separation energy of 0.975 MeV for the valence 
neutrons, that the nuclear breakup contributions could be substantial just 
around and above the grazing angle \cite{ban98}. The one-neutron separation 
energy in $^{14}$B is almost 1 MeV. Since the upper limits of detection angles
in the MSU data were around
or slightly above the grazing angle, it is possible that some contribution from
nuclear breakup effects is also present in the above data. This could be the
reason for the calculated results being smaller than the experimental ones.

We expect that breakup observables measured at very forward angles
(below the grazing angle) will not be affected by strong interactions and
also Coulomb-nuclear interference for this exotic nucleus. 
In Fig. 2, we have shown the calculated results for the $^{13}$B (ground 
state)$- n$ relative energy spectra for the same reaction considered above.
The angular integration for the scattering of the projectile-target relative
system in final channel has been performed up to the grazing angle of 4.4$^\circ$.
The AD and FRDWBA theories again give almost the same results. The
pure $s$-state energy spectra (solid lines) have a strong peak at very low 
relative energy around 0.4 MeV. This is very much characteristic of the
halo nuclei for breakup on a heavy target\cite{nak}. The $d$-state spectra 
(dashed lines) are almost two 
orders of magnitude smaller than the $s$-state spectra. We also show the
coherent sum of the two separate contributions weighted by the spectroscopic
factors of Ref.~\cite{war} by the long-dashed lines. This also has a prominent
peak around 400 keV relative energy. Since nuclear breakup effects could be
important at relative energies around and beyond 1 MeV \cite{ban99,dasso},
we suggest that measurements be done at small relative energies.

The exclusive neutron angular distribution in the one-neutron removal reaction
on a heavy target 
has been found to be very much forward peaked for the one-neutron halo nuclei 
\cite{ann}. This is a definite indication of the presence of a neutron halo 
structure. Therefore, we give predictions for the exclusive neutron angular 
distribution
for the same above reaction in Fig. 3. The angular integration with respect to
the core fragment has been done up to 4.4$^\circ$. The results for the 
pure $s$-state (solid lines) and coherent sum of $s$- and $d$-state 
contributions weighted by the same spectroscopic factors (long-dashed lines)
are almost the same within both the formalisms. The pure $d$-state results 
(dashed lines) are slightly different at very forward angles. Here also
the pure $d$-state contributions are two orders of magnitude smaller than
the pure $s$-state ones. But none of 
these distributions is quite forward peaked. Nevertheless, these results
together with the above predictions will prove useful in getting structure
informations about the ground state of $^{14}$B.

In conclusion, we have studied Coulomb breakup of the neutron rich exotic
nucleus $^{14}$B and compared our calculations with the recently available 
experimental data on a heavy target
in order to put constraint on the spectroscopic factors of its possible
ground state configurations. The breakup amplitude is
calculated within two approximate quantum mechanical theoretical models.
The adiabatic model
assumes that the important excitations of the projectile are to
the low-energy continuum so that they can be treated adiabatically. 
The distorted wave Born approximation method, on the other hand, assumes
that the breakup states are weakly coupled. Both
the methods permit finite-range treatment of the projectile vertex and
include initial and final state Coulomb interactions to all orders.
 
Using a single particle potential model for $^{14}$B, 
we find that calculations using the configuration $^{13}$B(${3\over 2}^-)\otimes
$2$s_{1/2}$ for the ground state of $^{14}$B with a spectroscopic factor 
$\approx$ 1 come closest to the MSU data on parallel momentum distribution of 
$^{13}$B (ground state) resulting from the fragmentation of $^{14}$B. This
is true within both the formalisms. This gives support to the existence of a 
one-neutron halo structure in $^{14}$B. The
small discrepancy between the magnitudes of the theoretical and experimental
cross sections could be attributed to the presence of nuclear
breakup contribution present in the kinematical domain covered by the data.
As far as Coulomb dissociation is concerned, it is suggested that further
measurements on breakup on heavy targets at extreme forward angles (below
the grazing angle) are necessary to derive very precise spectroscopic 
informations about the two possible configurations in the ground state of
$^{14}$B considered in this work. With this end in view, we have given
predictions for the exclusive neutron angular distribution and core-valence
relative energy spectrum in the one-neutron removal Coulomb breakup reaction
of $^{14}$B on a heavy target when the core remains in the ground state.

The authors would like to thank Prof. Jim Kolata for several helpful
correspondences and for providing the data shown in Fig. 1. Thanks are also
due to Prof. R. Shyam for his valuable suggestions.

\newpage
\begin{figure}
\begin{center}
\mbox{\epsfig{file=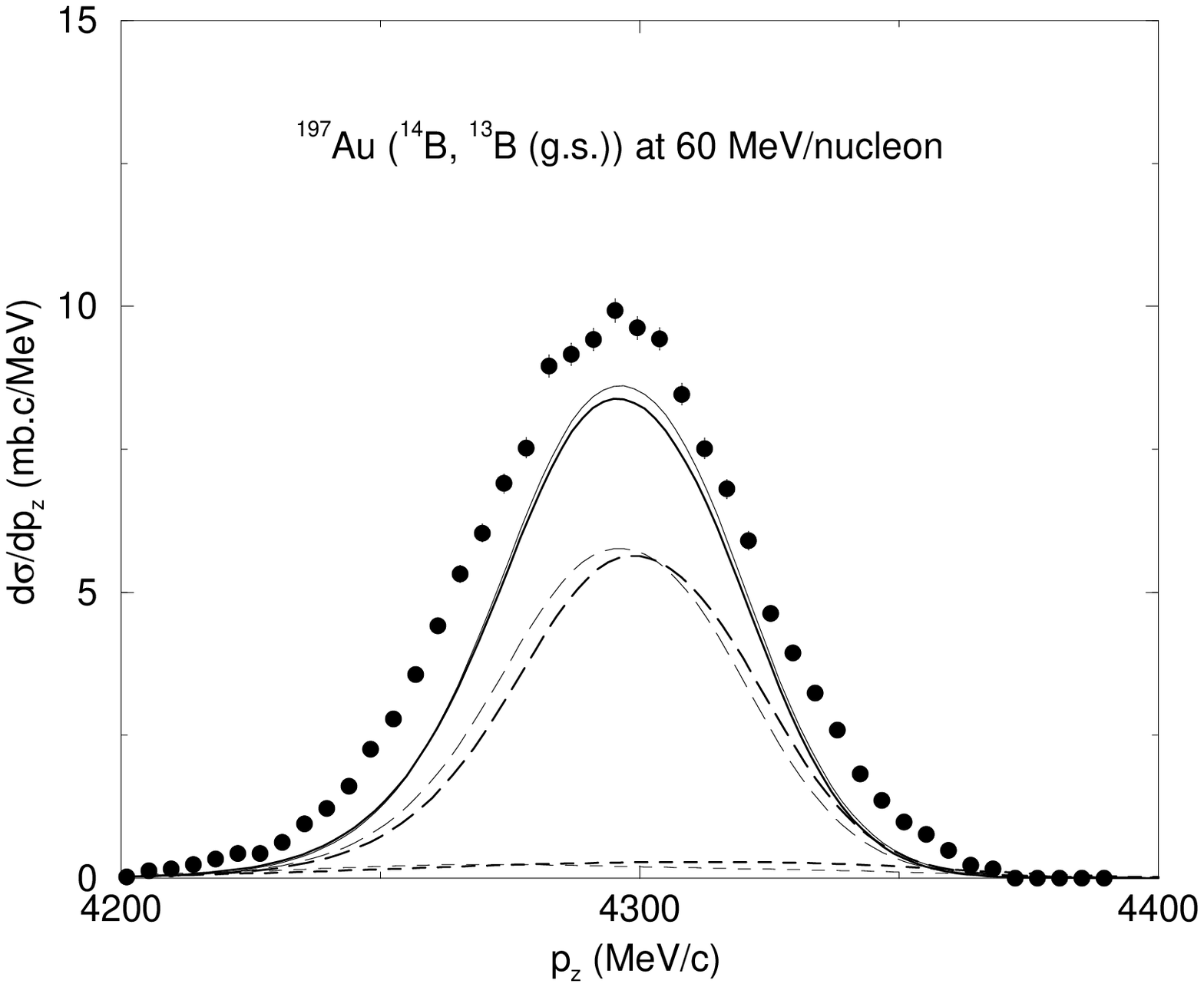,height=10cm}}
\end{center}
\caption{Calculated parallel momentum distributions of $^{13}$B (ground state)
for the Coulomb breakup of $^{14}$B on Au at 60 MeV/nucleon beam energy. The 
solid and dashed lines give pure $s$-state and pure $d$-state contributions
respectively, while the long-dashed curves are the results of coherent 
superposition of these two contributions weighted by spectroscopic factors 
given in Ref.~\protect\cite{war}. 
The thick and thin sets of lines are results of calculations of AD 
and FRDWBA theories respectively.
The data have been reported in Ref.~\protect\cite{gui}.}
\label{fig:figa}
\end{figure}

\begin{figure}
\begin{center}
\mbox{\epsfig{file=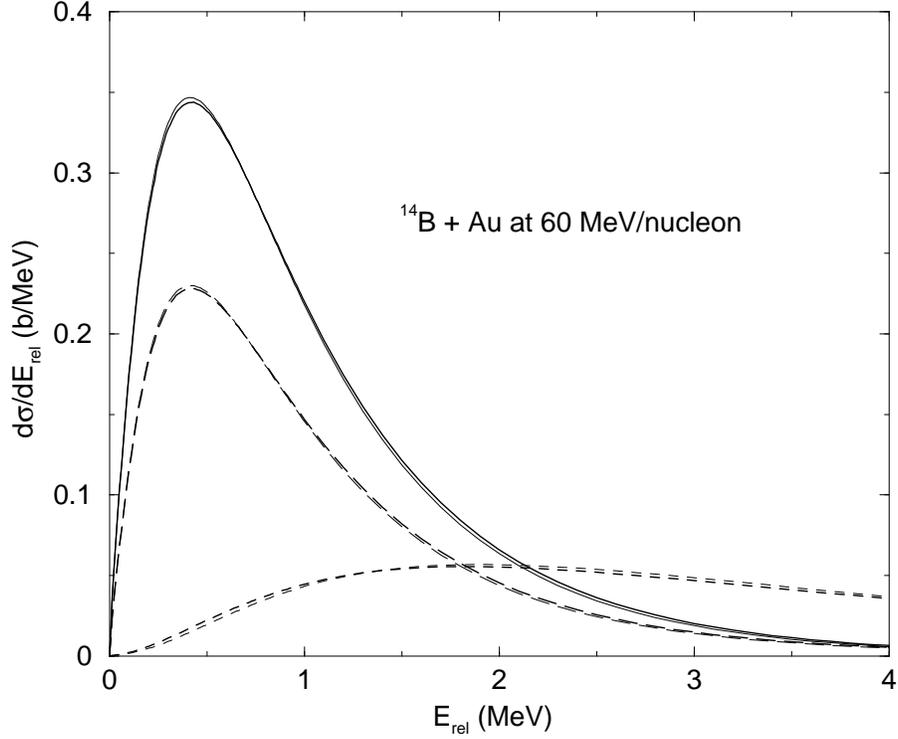,height=10cm}}
\end{center}
\caption{Calculated core (ground state) - valence neutron relative energy 
spectra for the Coulomb breakup of $^{14}$B on Au at 60 MeV/nucleon beam 
energy. The solid and dashed lines give pure $s$-state and (multiplied by 10)
pure $d$-state contributions respectively, while the long-dashed curves are the 
results of coherent superposition of these two contributions weighted by 
spectroscopic factors given in Ref.~\protect\cite{war}.
The thick and thin sets of lines are results of calculations of AD 
and FRDWBA theories respectively.
}
\label{fig:figb}
\end{figure}

\begin{figure}
\begin{center}
\mbox{\epsfig{file=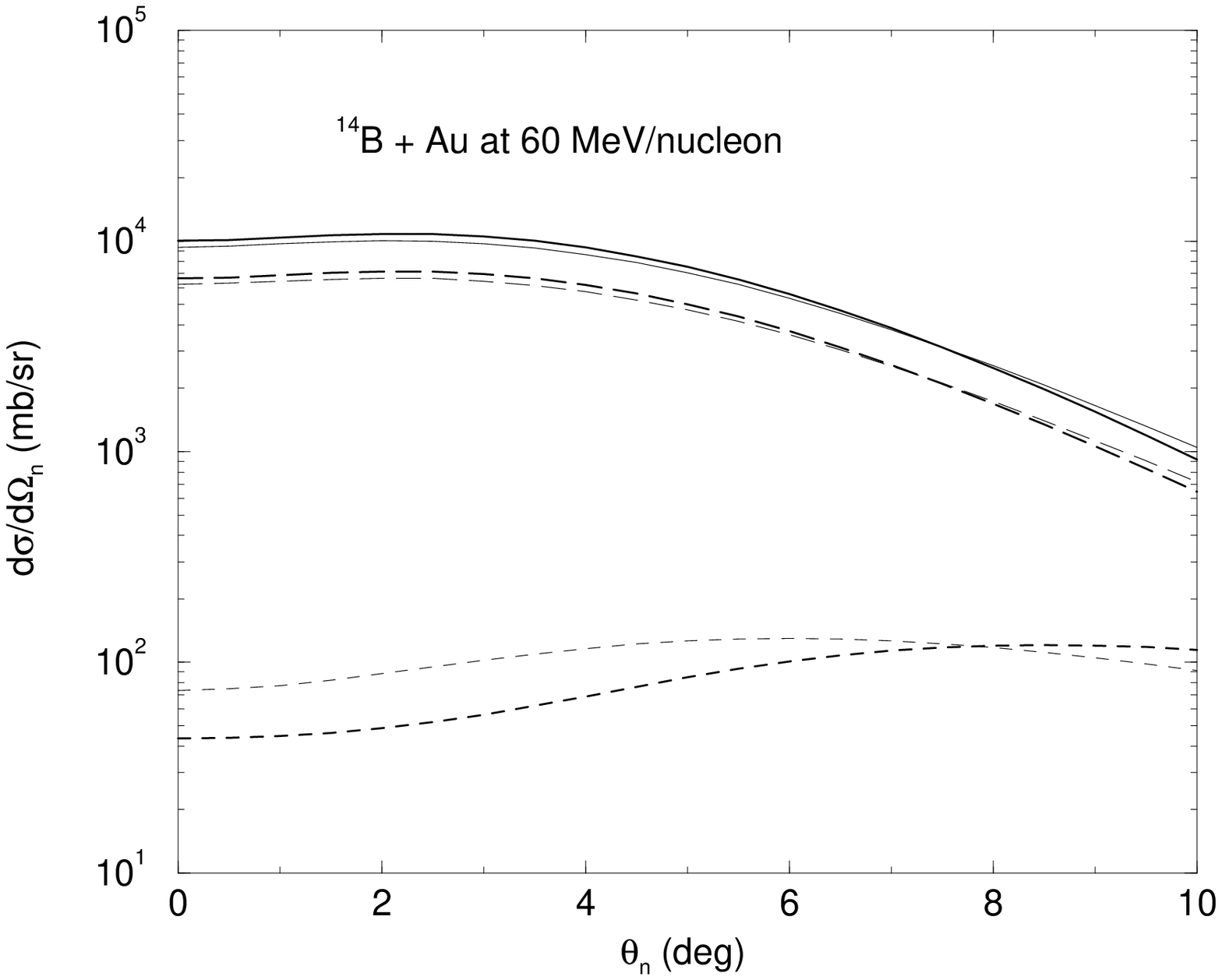,height=10cm}}
\end{center}
\caption{Calculated exclusive neutron angular distributions
for the Coulomb breakup of $^{14}$B on Au at 60 MeV/nucleon beam 
energy when the $^{13}$B core remains in the ground state. The solid and 
dashed lines give pure $s$-state and 
pure $d$-state contributions respectively, while the long-dashed curves are the 
results of coherent superposition of these two contributions weighted by 
spectroscopic factors given in Ref.~\protect\cite{war}.
The thick and thin sets of lines are results of calculations of AD 
and FRDWBA theories respectively.
}
\label{fig:figc}
\end{figure}

\begin{references}
\bibitem{han} P. G. Hansen, A. S. Jensen and B. Jonson, Ann. Rev. Nucl. Part.
Sci. {\bf 45}, 591 (1995).
\bibitem{baz} D. Bazin {\em et al.}, Phys. Rev. {\bf C57}, 2156 (1998).
\bibitem{nav} A. Navin {\em et al.}, Phys. Rev. Lett. {\bf 81}, 5089 (1998).
\bibitem{sai} M. G. Saint-Laurent {\em et al.}, Z. Phys. {\bf A332}, 457 (1989);
E. Liatard {\em et al.}, Europhys. Lett. {\bf 13}, 401 (1990).
\bibitem{vil} A. C. C. Villari {\em et al.}, Phys. Lett. {\bf B268}, 345 (1991).
\bibitem{ren} Zhongzhou Ren {\em et al.}, Z. Phys. {\bf A357}, 137 (1997).
\bibitem{sau} E. Sauvan {\em et al.}, LANL Preprint nucl-ex/0007013, submitted
to Phys. Lett. {\bf B}.
\bibitem{chatt} R. Chatterjee, P. Banerjee and R. Shyam, Nucl. Phys. {\bf A675},
477 (2000).
\bibitem {gui} V. Guimar$\widetilde{a}$es {\em et al.}, Phys. Rev. {\bf
C61}, 064609 (2000).
\bibitem {tos} J. A. Tostevin {\em et al.}, Phys. Letts. {\bf B424}, 219 (1998).
\bibitem{ban} P. Banerjee, I. J. Thompson and J. A. Tostevin, Phys. Rev.
 {\bf C58}, 1042 (1998).
\bibitem{ban99} P. Banerjee and R. Shyam, Phys. Rev. {\bf C61}, 047301 (2000).
\bibitem{nor} A. Nordsieck, Phys. Rev. {\bf 93}, 785 (1954).
\bibitem {war} E. K. Warburton and B. A. Brown, Phys. Rev. {\bf C46}, 923
(1992).
\bibitem{baz0} D. Bazin {\em et al.}, Phys. Rev. Lett. {\bf 74}, 3569 (1995). 
\bibitem{kol} J. J. Kolata, University of Notre Dame, private communications
(2000).
\bibitem{ban98} P. Banerjee, J. A. Tostevin, and I. J. Thompson, Phys. Rev.
 {\bf C58}, 1337 (1998).
\bibitem{nak} T. Nakamura {\em et al.}, Phys. Lett. {\bf B331}, 296 (1996);
Phys. Rev. Lett. {\bf 83}, 1112 (1999).
\bibitem{dasso} C. H. Dasso, S. M. Lenzi and A. Vitturi, Phys. Rev. {\bf C59},
539 (1999).
\bibitem{ann} R. Anne {\em et al.}, Nucl. Phys. {\bf A575}, 125 (1994);
\end{references}
\end{document}